\documentstyle[aps,prl,preprint]{revtex}

\begin{document}

\draft

\title{Non--linear supersymmetric $\sigma$--Model for Scalar Classical Waves}
\author{B. Elattari$^{1,2}$, V. Kagalovsky$^{1}$, and
H.A. Weidenm\"uller$^{1}$}
\address{$^1$Max-Planck-Institut f\"ur Kernphysik, 69029 Heidelberg,
Germany;\\
$^2$Universit\'e Choua\"\i b
Doukkali,  Facult\'e des Sciences, El Jadida,
Morocco}

\date{\today}

\maketitle

\begin{abstract}
  We derive a non--linear supersymmetric $\sigma$--model for 
  the transport of light (classical waves) through a disordered
  medium. We compare this model with the well--established non--linear
  $\sigma$--model for the transport of electrons (Schr\"odinger waves)
  and display similarities of and differences between both cases. We
  show that for weak disorder, both models are equivalent (have the
  same effective Lagrangean). This effective Lagrangean correctly
  reproduces the (different) Ward identities for Schr\"odinger waves
  and for classical waves.
\end{abstract}

\section{Introduction and Motivation}

In recent years, the transport of electrons and light through
disordered media has been studied intensely, and many interesting
effects have been observed and understood. Examples are universal
conductance fluctuations and weak localization for electrons, and
speckle patterns and enhanced backscattering for light. A thorough 
discussion may be found in Refs. \cite{review,reviewlight}. The universal
tool to deal with these and other phenomena in the case of electrons
has been Efetov's supersymmetric non-linear ${\bf{\sigma}}$--model (SUSIG) 
\cite{Efe83}. 
This model successfully describes not only the perturbative effects
mentioned above but also non--perturbative features like localization. 
It correctly accounts for both transport properties and spectral 
fluctuations. Thus, it is fair to say that SUSIG embodies the essence 
of electronic properties of disordered media. SUSIG has so far not been 
extended to the transmission of light through disordered media. In the 
present paper, we aim at filling this gap. Our motivation for this work
is the following. 

The transmision of light through disordered media is commomnly
described in terms of the scalar wave equation rather than a variant
of Maxwell's equations \cite{reviewlight}. The scalar wave equation differs
in a fundamental way from the Schr\"odinger equation for electrons,
see Sec. \ref{model}. By the same token, the Ward identities for both
equations differ substantially.  In trying to extend SUSIG to the
scalar wave equation, we probe the ability of the non--linear
${\bf \sigma}$--model to provide a universal description of wave propagation
in disordered systems described by different wave equations. It is of
interest to see in which way the difference in wave equations is
reflected in the effective Lagrangean of the resulting SUSIG. We will
show that SUSIG does apply to the scalar wave equation, and that
Efetov's effective Lagrangean is universal: It has the same form for
Schr\"odinger waves and for scalar waves. Using the replica trick,
John and Stephen \cite{john} derived a non--linear ${\bf \sigma}$--model for
classical waves. This derivation was confined, however, to waves at
fixed energy and thus bypassed the crucial issue of correlations
between amplitudes at {\it different} energies. The latter play the
central role in SUSIG.

The paper is organized as follows. In section \ref{model} we compare
the wave equations for electrons and classical waves and derive the
simplest variant of the Ward identities for both. Section \ref{sup}
describes the derivation of non--linear ${\bf \sigma}$--model for classical
waves in the supersymmetric formalism. We compare our result with the
analogous expression in Efetov's work \cite{Efe83}. Section \ref{war}
is devoted to the derivation of the Ward identities within a
supersymmetric formalism for classical waves.  Our conclusions are
presented in section \ref{con}.

\section{Wave Equations}
\label{model} 

We put the mass $m$ of the electron and Planck's constant $\hbar$
equal to unity. The Schr\"odinger equation for a noninteracting
electron,
\begin{equation}
(-\frac{1}{2}\Delta+V(r))\phi =E\phi \ ,
\label{shr}
\end{equation}
contains the random potential $V(r)$ which describes impurity scattering.
The propagation of light is described by the classical wave equation
\begin{equation}
(-\Delta +k^2\delta\epsilon ({\bf r}))\phi =k^2\phi \ .
\label{clw}
\end{equation}
Here, $k = \omega / c$ is the wave number. We have decomposed the
space--dependent dielectric constant $\epsilon(r) = 1 -
\delta \epsilon({\bf r})$ into a space--independent background term
(which we put equal to $1$) and a fluctuating part $\delta
\epsilon({\bf r})$. We assume
that $\delta \epsilon$ is a Gausian random process with vanishing
first moment and a second moment given by
\begin{equation}
<\delta\epsilon ({\bf r_1})\delta\epsilon ({\bf r_2})>=
\frac{4\pi}{\ell k^{d+1}}\delta ({\bf r_1}-{\bf r_2}) \ ,
\label{diel}
\end{equation}
where $l$ is the elastic mean free path and $d$ is the dimension of
the system. We have written Eq. (\ref{diel}) in complete analogy to
the case of electrons.

We compare Eqs. (\ref{shr}) and (\ref{clw}). Aside from a factor two,
the quantity $k^2$ corresponds formally to the energy
$E$. However, $k^2$ is always positive, in contradistinction
to $E$. The main difference between Eqs. (\ref{shr}) and (\ref{clw})
lies in the energy dependence of the random potential in the classical
case. While $V$ is independent of energy, the analogous term $\delta
\epsilon$ is not. This difference is also reflected in different Ward
identities which relate averaged one-- and two--point functions. For
electrons, the retarded (advanced) Green function
\begin{equation}
G_e^{\pm}=(E^{\pm}\pm z+\Delta /2-V)^{-1} \ 
\label{gre12}
\end{equation}
is taken at energy $E^{+}+z=E+z+i\eta$ ($E^{-}-z=E-z-i\eta$),
respectively. We immediately find
\begin{equation}
<G_e^{+}>-<G_e^{-}>=-2(z+i\eta)<G_e^{+}G_e^{-}> \ ,
\label{wae}
\end{equation}
where the angular brackets stand for the ensemble average.
In complete analogy, we define the Green functions for classical waves by
\begin{equation}
G_c^{\pm}=\left[\left( k_{0}^{2\pm} \pm \frac{\Delta k^2}{2}\right)
 + \Delta - \left( k_0^2
\pm \frac{\Delta k^2}{2}\right) \delta \epsilon\right]^{-1} \ .
\label{grc}
\end{equation}
The Ward identity reads
\begin{equation}
<G_c^{+}>-<G_c^{-}>=-2\left(\frac{\Delta
k^2}{2}+i\eta\right) <G_c^{+}G_c^{-}>+\Delta k^2 <G_c^{+} \delta\epsilon
G_c^{-}> \ .
\label{wac}
\end{equation}
Because of the frequency dependence of the impurity term in the
classical case, these two Ward identities differ in form. They
actually also indicate different conservation laws: Particle
conservation for electrons, and energy conservation for classical
waves. The Ward identity for the classical case will serve as a check
of our supersymmetric formalism: In section \ref{war} we derive it
from the non--linear ${\bf \sigma}$--model.

\section{Non--linear sigma model}
\label{sup}

We derive the non--linear ${\bf \sigma}$--model for the simplest
non--trivial case, the ensemble average of a product of an advanced
and a retarded Green function. We use the notations and definitions of
ref. \cite{Ver85}. The advanced and retarded Green functions can be
written as integrals over supervectors
\begin{equation}
G^{\pm}({\bf y_{1}},{\bf y_{2}},k^{2})=\mp\frac{i}{2}\int {\cal D}[\Psi
]\Psi_{\nu}({\bf y_{1}})\Psi_{\nu}^{\dagger}({\bf y_{2}})
\exp [{\cal L}(\Psi )] \ ,
\label{gnn}
\end{equation}
where we have omitted the index $c$ for the Green's function of 
classical waves, and the Lagrangean is given by 
\begin{equation}
{\cal L}=\frac{1}{2}i\int d^d{\bf y}\left[ \Psi^{\dagger}({\bf y})
\left(\pm\left( k^2+\Delta -\delta\epsilon 
({\bf y})k^2\right)+i\eta \right)\Psi ({\bf y})\right] \ . 
\label{lag}
\end{equation}
The quantities $\Psi({\bf x})$ are supervectors defined by
\begin{equation}
\Psi({\bf x})^{\dagger}=(S^1({\bf x}),S^2({\bf x}),-\chi({\bf x}) \ ,
\chi^{*}({\bf x})) \ .
\label{vec}
\end{equation}
The quantities $S$ are ordinary real integration variables, and the
$\chi$'s anticommute. We introduce a source term $J({\bf
y})=$diag$(j({\bf y}),0,0,0)$ in graded
space and introduce the generating functional
\begin{equation}
Z^{\pm}(k^2,J)=\int {\cal D}[\Psi ] \exp \left[ {\cal L} + \frac{1}{2} i 
\int d^d{\bf y} (\Psi^{\dagger}({\bf y}) J({\bf y}) \Psi ({\bf y})) \right] \ .
\label{gen}
\end{equation}
This functional generates the Green function at point ${\bf y_1}=
{\bf y_2}$, which is sufficient, because in the present section we are
interested only in
the effective action. In section IV we show how to
generate the Green function with different space point arguments. 
The Green function is given as functional 
derivative of the
generating functional with respect to $J$ at $J = 0$,
\begin{equation}
G^{\pm}({\bf y},{\bf y})=\mp\frac{\partial Z^{\pm}}{\partial j({\bf
    y})} \ .
\label{gre}
\end{equation}
We use this expression to calculate the average of the product of a
retarded and an advanced Green function taken at different
frequencies, $<G^{+} G^{-}>$ (the two--point function). This quantity
plays an important role in describing average properties of random
systems, such as the level--level correlation function, the
distribution function of the transmission, etc. It serves as an
example. Except for the dimension of the $Q$--matrices appearing
below, and except for the dependence on additional frequency variables,
the average $2k$--point function for any positive integer $k$ is
governed by an effective Lagrangean of the same type.

The generating functional $Z$ for the two--point function is given by
\begin{eqnarray}
&&Z(k^2,\Delta k^2,J)=
\nonumber \\
&&\int {\cal D}[\Psi ] \exp\left[
\frac{1}{2}i\int d^d{\bf y}\left[
\Psi^{\dagger}({\bf y})L^{1/2}\left( k_{0}^{2}+
\frac{\Delta k^2}{2}{\rm L}+\Delta 
-\delta\epsilon ({\bf y})\left(k_{0}^{2}+\frac{\Delta k^2}{2}L\right)
\right.\right.\right.
\nonumber \\
&&\left.\left.\left. +i\eta {\rm L}+ J({\bf y}) \right)L^{1/2}\Psi ({\bf y})
\right]\right] \ ,
\label{gen1} 
\end{eqnarray}
where $L=$diag$(1,1,1,1,-1,-1,-1,-1)$, $\Psi$ are supervectors with
$8$ components, and $J$ is an $8\times 8$ matrix. All quantities are
given in ``advanced--retarded'' notation (see
Ref. \onlinecite{Ver85}). Averaging over the Gaussian distribution of
$\delta \epsilon$, we obtain the Lagrangean
\begin{eqnarray}
&&{\cal L}=\frac{1}{2}i\int d^d{\bf y}\left[ \Psi^{\dagger}({\bf
  y})L^{1/2}\left( k_{0}^{2}+\frac{\Delta k^2}{2}L+
\Delta+i\eta L \right)\Psi ({\bf y})L^{1/2}-
\right.
\nonumber \\
&&\left.\frac
{\pi}{2k_{0}^{d-3}\ell}
\left(\Psi^{\dagger}({\bf y})L^{1/2}(1+\frac {\Delta k^2}{2k_{0}^{2}}L)
{\rm L}^{1/2}\Psi({\bf y})\right)^2\right] \ .
\label{lag1}
\end{eqnarray}
Using the Hubbard--Stratonovich transformation in the usual way and
integrating over the vectors $\Psi$, we obtain the
following form of the generating functional.
\begin{eqnarray}
&&\overline {Z}=\int {\cal D}Q\exp\left[\int d^d{\bf y}\left[
-\frac{\pi\nu}{8\tau}{\rm trg}Q^2+\frac{1}{2}{\rm trg}
\log\left( k_{0}^{2}+\frac{\Delta k^2}{2}{\rm L}+
\Delta +i\eta {\rm L}+J({\bf y})-
\right. \right. \right.
\nonumber \\
&&\left. \left. \left.
\frac{1}{2\tau}Q\left( 1+\frac{\Delta k^2}{2k_{0}^{2}}{\rm
  L}\right)\right) \right]\right] \ .
\label{action}
\end{eqnarray}
Here $\nu$ is the density of states per unit of $k_{0}^{2}$ and per
unit of volume, and $\tau = k_{0}^{d-3}\ell/(2\pi ^2\nu )$ formally
corresponds to Efetov's mean free time \cite{Efe83}. We have
introduced these quantities in Eq. (\ref{action}) in order to
facilitate the direct comparison to Efetov's expression for
electrons. The term $(1 / 2 \tau) Q {\rm L} ( \Delta k^2 / 2 k_0^2)$
is due to the frequency dependence of the ``scattering potential''
$k^2 \delta \epsilon$ in Eq. (\ref{clw}). Comparing Eq. (\ref{action})
with the corresponding expression in Efetov's work \cite{Efe83}, we
identify (modulo factors of two) $\epsilon_0 k_0^2$ with the sum
energy and $\epsilon_0 \Delta k^2$ with the energy difference and find
that the two expressions differ by the term $(1 / 2 \tau) Q {\rm L} (
\Delta k^2 / 2 k_0^2)$. 

To evaluate Eq. (\ref{action}), we use the saddle--point approximation. 
This is justified if $\tau \ll \rho$, the mean level density. Varying the
Lagrangean in Eq. (\ref{action}) with respect to $Q$ and neglecting
terms proportional to $\Delta k^2$ and source terms, we obtain the
standard saddle--point equation
\begin{equation}
Q=\frac{1}{\pi\rho}{\rm tr}\left[ k_{0}^{2}+\Delta-
\frac{1}{2\tau}Q\right]^{-1} \ .
\label{sad}
\end{equation}
This is the same equation as in the case of electrons. As in that
case, the condition $k_0\ell\gg 1$ (weak disorder) yields $Q=i{\rm L}$
as a solution of the saddle--point equation. The weak disorder condition
also implies, however, that the term $(1 / 2 \tau) Q {\rm L} (\Delta
k^2 / 2 k_0^2)$ in Eq. (\ref{action}) can be neglected. This is the
case for sufficiently large $k_0$. Then, {\it there is no difference
between the non--linear ${\bf \sigma}$--models for Schr\"odinger waves and
for classical waves}. This statement is the central result of our work. 
It obviously extends to the generating functionals of all higher 
correlation functions and, thus, applies universally. 

The actual differences between the two theories are due to the
different forms of the source terms. In the next Section, we show this
in the case of the Ward identities.

\section{Ward Identity}
\label{war}

In Appendix G of Ref. \onlinecite{Ver85}, it was shown how a Ward
identity can be derived in the context of SUSIG. We use that method to
check the Ward identity, Eq. (\ref{wac}), for classical waves, using
essentially the generating functional derived in the previous
Section. With slight modifications, our calculation also applies to
the case of electrons. We first 
show how the new source terms emerge, when we introduce a new 
generating functional for the r.h.s. of 
Eq. (\ref{wac}). We use the coordinate representation 
\begin{equation}
<{\bf r}|G^{+}(1-\delta\epsilon )G^{-}|{\bf r'}>=
\int d^d{\bf x}G^{+}({\bf x},{\bf r})(1-\delta\epsilon ({\bf x}))G^{-}
({\bf r'},{\bf x}) \ .
\label{coor}
\end{equation}
The generating functional
\begin{eqnarray}
&&Z_1(k^2,\Delta k^2,J_1)=
\nonumber \\
&&-\int {\cal D}[\Psi_1] \exp \Bigg[ \frac{i}{2}\int
d^d{\bf y}
\Psi^{\dagger}_{1}({\bf y})\left[k_{0}^{2}+
\frac{\Delta k^2}{2}+\Delta 
-\delta\epsilon ({\bf y})\left(k_{0}^{2}+\frac{\Delta
  k^2}{2}\right)+i\eta\right]\Psi_1 ({\bf y'})+
\nonumber \\ 
&&\frac{i}{2}\int\int d^d{\bf y}d^d{\bf y'}\Psi^{\dagger}_{1}({\bf y})
J_1({\bf y},{\bf y'})\Psi_1({\bf y'}) \Bigg] \ , 
\label{genfir} 
\end{eqnarray}
produces the retarded Green's function on the r.h.s. of Eq. (\ref{coor}):
\begin{equation}
G^{+}({\bf x},{\bf r})=\frac{\partial Z_1}{\partial 
j_1({\bf r},{\bf x})} \ ,
\label{gre1}
\end{equation}
where the source term  $J_1={\rm diag}(j_1,0,0,0)$ is put equal to
zero after taking the derivative, while $\Psi_1$ is a supervector with $4$ 
components. We also introduce another generating functional
\begin{eqnarray}
&&Z_2(k^2,\Delta k^2,{\tilde J})=
\nonumber \\
&&-\int {\cal D}[\Psi_2] \exp [-\frac{i}{2}\int
d^d{\bf y}
\Psi^{\dagger}_{2}({\bf y})\left[ k_{0}^{2}-
\frac{\Delta k^2}{2}+\Delta 
-\delta\epsilon ({\bf y})\left(k_{0}^{2}-\frac{\Delta
  k^2}{2}\right)-i\eta\right]\Psi_2 ({\bf y'})
\nonumber \\ 
&&-\frac{i}{2}\int\int d^d{\bf y}d^d{\bf y'}\Psi^{\dagger}_{2}({\bf y})
(1-\delta\epsilon ({\bf y})){\tilde J}({\bf y},{\bf y'})\Psi_2({\bf
  y'})] \ .
\label{gensec} 
\end{eqnarray}
Then, immediately
\begin{equation}
(1-\delta\epsilon ({\bf x}))G^{-}({\bf r'},{\bf x})=
\frac{\partial Z_2}{\partial j_2({\bf x},{\bf r'})} \ ,
\label{gre2}
\end{equation}
Taking the product of Eqs. (\ref{genfir},\ref{gensec}), 
we obtain a generating functional
\begin{equation}
Z_f=\int {\cal D}[\Psi ] \exp\left[\frac{i}{2}{\cal L}_f\right] \ ,
\label{zfin}
\end{equation}
where the action is given by
\begin{eqnarray}
&&{\cal L}_f=\int\int d^d{\bf y}d^d{\bf y'}\Psi^{\dagger}({\bf y}){\rm L}^{1/2}
\Big[\Big(k_{0}^{2}+
\frac{\Delta k^2}{2}{\rm L}+\Delta 
-\delta\epsilon ({\bf y})\big(k_{0}^{2}+\frac{\Delta
  k^2}{2}{\rm L})+i\eta {\rm L}\Big)\delta ({\bf y}-{\bf y'})+
\nonumber \\
&& 
J({\bf y},{\bf y'})
-\delta\epsilon ({\bf y'}){\tilde J}({\bf y},{\bf y'})\Big]{\rm L}^{1/2}
\Psi ({\bf y'}) \ ,
\label{finact}
\end{eqnarray}
with ${\tilde J}={\rm diag}(0,0,0,0,j_2,0,0,0)$, 
$J={\rm diag}(j_1,0,0,0,j_2,0,0,0)$. The second partial derivatives 
of $Z_f$ produce the
integrands on the r.h.s. of Eq. (\ref{wac}). The additional
source term ${\tilde J}({\bf y},{\bf y'})$ represents the important
difference to 
the electron case.
Averaging of the term containing the random part of the dielectric 
constant $\delta\epsilon ({\bf y})$ leads to
\begin{equation}
\exp \left[ -\frac{1}{16\pi\nu\tau k_{0}^{4}}\int\int\int 
 d^d{\bf y}d^d{\bf y'}d^d{\bf y'_1} 
\Psi^{\dagger}({\bf y}){\rm L}^{1/2}A({\bf y},{\bf y'}){\rm L}^{1/2}
\Psi ({\bf y'})
\Psi^{\dagger}({\bf y}){\rm L}^{1/2}A({\bf y},{\bf y'_1}){\rm L}^{1/2}
\Psi ({\bf y'_1})\right]  
\label{action01}
\end{equation}
where
\begin{equation}
A({\bf y},{\bf y'})=\left[k_{0}^{2}+\frac{\Delta k^2}{2}{\rm L}\right]
\delta ({\bf y}-{\bf y'})+{\tilde J}({\bf y},{\bf y'}) \ .
\label{aaa}
\end{equation}
To perform the Hubbard--Stratonovich transformation
we introduce a supervector
\begin{equation}
F({\bf y})=\int d^d{\bf y'}A({\bf y},{\bf y'}){\rm L}^{1/2}\Psi ({\bf
  y'}) \ ,
\label{fff}
\end{equation}
after which we can rewrite the expression in Eq. (\ref{action01}) as
\begin{equation}
\exp \left[ -\frac{1}{16\pi\nu\tau k_{0}^{4}}\int 
d^d{\bf y}[\Psi^{\dagger}({\bf y}){\rm L}^{1/2}F({\bf y})]^2\right] \ .
\label{action02}
\end{equation}
Using the Hubbard--Stratonovich transformation and keeping only diffusive
modes, we obtain the average of 
Eq. (\ref{zfin}) 
\begin{equation}
\overline {Z_f}=\int {\cal D}[\Psi ]\exp\left[-\int d^d{\bf y}
\frac{\pi\nu}{8\tau}{\rm trg}Q^2- 
\frac{1}{2}\log{\rm Detg}B(Q)\right]\equiv\int {\cal D}[Q]
\exp [-{\cal L}_i(Q)] \ ,
\label{action1}
\end{equation}
where Detg means determinant over real and graded spaces and we define
a matrix
\begin{equation}
B(Q)=\left(\left[k_{0}^{2}+\frac{\Delta k^2}{2}{\rm L}+
\Delta +i\eta {\rm L}\right]\delta ({\bf y}-{\bf y'})
+J({\bf y},{\bf y'})-\frac{1}{2\tau k_{0}^{2}}Q(\bf y )A({\bf y},{\bf y'})
\right) \ .
\label{action11}
\end{equation}

For maximum compactness we allow $J$ and ${\tilde J}$ to be general 
symmetric $8\times 8$ matrices.
This is permissible, because we never use the 
particular form of the source term in the derivation in the previous
section.  
The saddle--point equation for this action is the same as before. 
Following the formalism developed in Ref. \onlinecite{Ver85} we apply 
the transformation $Q\rightarrow (1+\delta T)^{-1}Q(1+\delta T)$
 (we preserve the notation of ref. \onlinecite{Ver85}), changing the 
action in Eq. (\ref{action1}) into 
\begin{eqnarray}
&&{\cal L}_i+\frac{1}{2}{\rm Trg}B^{-1}(Q)
\left(\left[\delta T,(\frac{\Delta k^2}{2}{\rm L}
+i\eta {\rm L})\delta ({\bf y}-{\bf y'})
+J({\bf y},{\bf y'})\right]-
\frac{1}{2\tau}Q({\bf y})\times
\right.
\nonumber \\
&&\left.\left[\delta T,
\frac{\Delta k^2}{2k_{0}^{2}}{\rm
  L}\delta ({\bf y}-{\bf y'})
+
\frac{{\tilde J}({\bf y},{\bf y'})}{k_{0}^{2}}\right]\right) \ ,
\label{action2}
\end{eqnarray}
where Trg is the trace in both, real and graded, spaces.
A transformation of integration variables leaves $\overline {Z_f}$
invariant. Therefore, terms linear in $\delta T$ in the expression
of $\overline {Z_f}$ must vanish, which leads to the following
equation
\begin{eqnarray}
&&\int {\cal D}[Q]\exp (-{\cal L}_i){\rm Trg}B^{-1}(Q)
\bigg[\delta T,J\bigg]
\nonumber \\
&&+\int {\cal D}[Q]\exp (-{\cal L}_i){\rm Trg}B^{-1}(Q)
\Bigg[\delta T,(\frac{\Delta k^2}{2}
+i\eta ){\rm L}\delta ({\bf y}-{\bf y'}) \Bigg]
\nonumber \\
&&-\frac{1}{2\tau k_{0}^{2}}
\int {\cal D}[Q]\exp (-{\cal L}_i){\rm Trg}B^{-1}(Q)Q
\Bigg[\delta T,((\frac{\Delta k^2}{2}{\rm L}\delta ({\bf y}-{\bf y'})+
{\tilde J}({\bf y},{\bf y'}))\Bigg]=0
\label{nul}
\end{eqnarray}
We will    
consider each of these terms in detail. The first term is
\begin{eqnarray}
&&\int {\cal D}[Q]\exp (-{\cal L}_i){\rm Trg}[B^{-1}
({\bf y},{\bf y'})]^{\alpha\beta}
\bigg[\delta T,J\bigg]_{{\bf y},{\bf y'}}^{\alpha\beta}=
\nonumber \\
&&\sum_{\alpha\beta}\int\int d^d{\bf y}d^d{\bf y'}\int {\cal D}[Q] 
\exp (-{\cal L}_i)B^{-1}_{\alpha\beta}(Q)\bigg[\delta T,J\bigg]^{\beta\alpha}=
\nonumber \\ 
&&\sum_{kk'}\int\int d^d{\bf y}d^d{\bf y'}
\frac{\partial\overline{Z_f}(k_{0}^{2},\Delta
  k^2,J)}
{\partial
  J_{kk'}^{(1,1)}({\bf y'},{\bf y})}\bigg[\delta T(1,2)J(2,1)-
J(1,2)\delta T(2,1)\bigg]^{kk'}_{{\bf y'},{\bf y}}
\nonumber \\[0.3cm]
&&+\sum_{kk'}\int\int d^d{\bf y'}d^d{\bf y}
\frac{\partial\overline{Z_f}(k_{0}^{2},\Delta
  k^2,J)}
{\partial
  J_{kk'}^{(2,2)}({\bf y'},{\bf y})}\bigg[\delta T(2,1)J(1,2)-
J(2,1)\delta T(1,2)\bigg]^{kk'}_{{\bf y'},{\bf y}}+\ldots,
\label{per}
\end{eqnarray}
where we use block notation as in Ref. \onlinecite{Ver85}.
 The dots represent terms containing $J(1,1)$ and $J(2,2)$ 
which do not contribute to the final result. Using the explicit 
expressions $\delta T_{kk'}^{1,2}=~i\delta_{kk_0}\delta_{k'k_0^{'}}$,
$\delta T_{kk'}^{2,1}=-i\delta_{kk_{0}^{'}}\delta_{k'k_0}$,
the first term can be written as
\begin{eqnarray}
&&i\sum_{k'}\int\int d^d{\bf y}d^d{\bf y'}
\frac{\partial\overline{Z_f}(k_{0}^{2},\Delta
  k^2,J)}
{\partial
  J_{k_0k'}^{(1,1)}({\bf y'},{\bf y})} J_{k_{0}^{'}k'}^{(2,1)}({\bf
  y'},
{\bf y}) 
\nonumber \\[0.3cm]
&&+i\sum_{k}\int\int d^d{\bf y}d^d{\bf y'}
\frac{\partial\overline{Z_f}(k_{0}^{2},\Delta
  k^2,J)}
{\partial
  J_{kk_{0}}^{(1,1)}({\bf y'},{\bf y})} 
J_{kk_{0}^{'}}^{(1,2)}({\bf y'},{\bf y})-
\nonumber \\[0.3cm]
&&i\sum_{k'}\int\int d^d{\bf y}d^d{\bf y'}
\frac{\partial\overline{Z_f}(k_{0}^{2},\Delta
  k^2,J)}
{\partial
  J_{k_{0}^{'}k'}^{(2,2)}({\bf y'},{\bf y})} J_{k_0k'}^{(1,2)}({\bf
  y'},
{\bf y}) 
\nonumber \\[0.3cm]
&&-i\sum_{k}\int\int d^d{\bf y}d^d{\bf y'}
\frac{\partial\overline{Z_f}(k_{0}^{2},\Delta
  k^2,J)}
{\partial
  J_{kk_{0}^{'}}^{(2,2)}({\bf y'},{\bf y})} 
J_{kk_{0}}^{(2,1)}({\bf y'},{\bf y})+\ldots
\label{vtor}
\end{eqnarray}
Now taking the derivative with respect to 
$J_{11}^{(1,2)}({\bf x},{\bf x'})$, we finallly obtain
\begin{equation}
2i\frac{\partial\overline{Z_f}(k_{0}^{2},\Delta
  k^2,J)}
{\partial
  J_{1k_{0}}^{(1,1)}({\bf x},{\bf x'})}\Big|_{J,{\tilde J}=0}\delta _{1k_{0}^{'}}-
2i\frac{\partial\overline{Z_f}(k_{0}^{2},\Delta
  k^2,J)}
{\partial
  J_{k_{0}^{'}1}^{(2,2)}({\bf x},{\bf x'})}
\Big|_{J,{\tilde J}=0}\delta _{1k_0} \ .
\label{tre}
\end{equation}
We consider now the second term in 
Eq. (\ref{nul})
\begin{eqnarray}
&&\int {\cal D}[Q]\exp (-{\cal L}_i){\rm Trg}B^{-1}(Q) 
[\delta T,(\frac{\Delta k^2}{2}{\rm L}
+i\eta {\rm L})\delta ({\bf y}-{\bf y'})]=
\nonumber \\[0.3cm]
&&-2(\frac{\Delta k^2}{2}
+i\eta )\sum_{kk'}\int\int d^d{\bf y}d^d{\bf y'}
\frac{\partial\overline{Z_f}(k_{0}^{2},\Delta
  k^2,J)}
{\partial
  J_{kk'}^{(1,2)}({\bf y'},{\bf y})}\delta T_{kk'}(1,2)
\delta ({\bf y}-{\bf y'})
\nonumber \\[0.3cm]
&&+2(\frac{\Delta k^2}{2}
+i\eta )\sum_{kk'}\int\int d^d{\bf y}d^d{\bf y'}
\frac{\partial\overline{Z_f}(k_{0}^{2},\Delta
  k^2,J)}
{\partial
  J_{kk'}^{(2,1)}({\bf y'},{\bf y})}\delta T_{kk'}(2,1)
\delta ({\bf y}-{\bf y'})=
\nonumber \\[0.3cm]
&&-2i(\frac{\Delta k^2}{2}
+i\eta )\int d^d{\bf y}\frac{\partial\overline{Z_f}
(k_{0}^{2},\Delta  k^2,J)}
{\partial
  J_{k_0k_{0}^{'}}^{(1,2)}({\bf y},{\bf y})}-
2i(\frac{\Delta k^2}{2}
+i\eta )\int d^d{\bf y}\frac{\partial\overline{Z_f}
(k_{0}^{2},\Delta  k^2,J)}
{\partial
  J_{k_{0}^{'}k_{0}}^{(2,1)}({\bf y},{\bf y})} \ .
\label{chet}
\end{eqnarray}
The derivative with respect to 
$J_{11}^{(1,2)}({\bf x},{\bf x'})$ leads to 
\begin{eqnarray}
-2i(\frac{\Delta k^2}{2}+
i\eta )\bigg(\int d^d{\bf y}\frac{\partial^2\overline{Z_f}
(k_{0}^{2},\Delta  k^2,J)}
{\partial
  J_{k_0k_{0}^{'}}^{(1,2)}({\bf y},{\bf y})\partial
  J_{11}^{(1,2)}({\bf x},{\bf x'})}\Big|_{J,{\tilde J}=0}+
\nonumber \\
\int d^d{\bf y}\frac{\partial^2\overline{Z_f}
(k_{0}^{2},\Delta  k^2,J)}
{\partial
  J_{k_{0}^{'}k_0}^{(2,1)}({\bf y},{\bf y})\partial
  J_{11}^{(1,2)}({\bf x},{\bf x'})}\Big|_{J,{\tilde J}=0}\bigg) \ .
\label{pyt}
\end{eqnarray}
We expand the remaining term in Eq. (\ref{nul})
\begin{eqnarray}
&&-\frac{1}{2\tau k_{0}^{2}}\int {\cal D}[Q]\exp (-{\cal L}_i)
{\rm Trg}B^{-1}(Q)Q[\delta T,
(\frac{\Delta k^2}{2}{\rm L}\delta ({\bf y}-{\bf y'})+
{\tilde J}({\bf y},{\bf y'})]=
\nonumber \\[0.1cm]
&&\sum_{\alpha\beta}\int d^d{\bf y}\frac{\partial\overline{Z_f}
(k_{0}^{2},\Delta  k^2,J)}
{\partial
  {\tilde J}^{\alpha\beta}({\bf y},{\bf y})}\left[\delta T,\frac{\Delta k^2}{2}
{\rm L}\right]^{\alpha\beta}_{{\bf y},{\bf y'}}=
\nonumber \\[0.1cm] 
&&-i\Delta k^2
\int d^d{\bf y}\frac{\partial\overline{Z_f}
 (k_{0}^{2},\Delta  k^2,J)}
 {\partial{\tilde J}_{k_0 k_{0}^{'}}^{(1,2)}({\bf y},{\bf y})}
 -i\Delta k^2
\int d^d{\bf y}\frac{\partial\overline{Z_f}
(k_{0}^{2},\Delta  k^2,J)}
{\partial
  {\tilde J}_{k_{0}^{'}k_0}^{(2,1)}({\bf y},{\bf y})} \ .
\label{she}
\end{eqnarray}
The term proportional to ${\tilde J}$ is omitted, because it does not
contribute to the final result. 
Taking the derivative with respect to $J^{(1,2)}_{1,1}$, we find
\begin{eqnarray}
 -i\Delta k^2
\int d^d{\bf y}\frac{\partial^2\overline{Z_f}
(k_{0}^{2},\Delta  k^2,J)}
{\partial
  {\tilde J}_{k_0 k_{0}^{'}}^{(1,2)}({\bf y},{\bf y})
J_{11}^{(1,2)}({\bf x},{\bf x'})}\Big|_{J,{\tilde J}=0} 
\nonumber \\
-i\Delta k^2
\int d^d{\bf y}\frac{\partial^2\overline{Z_f}
(k_{0}^{2},\Delta  k^2,J)}
{\partial
  {\tilde J}_{k_{0}^{'}k_0}^{(2,1)}({\bf y},{\bf y})
J_{11}^{(1,2)}({\bf x},{\bf x'})}\Big|_{J,{\tilde J}=0} \ .
\label{sem}
\end{eqnarray}
In conclusion, the requirement that the term linear in $\delta T$ in the
expansion of $\overline{ Z_f}$ vanishes, entails the following equation :
\begin{eqnarray}
&&2\frac{\partial\overline{Z_f}(k_{0}^{2},\Delta
  k^2,J)}
{\partial
  J_{1k_{0}}^{(1,1)}({\bf x},{\bf x'})}\Big|_{J,{\tilde J}=0}\delta _{1k_{0}^{'}}-
2\frac{\partial\overline{Z_f}(k_{0}^{2},\Delta
  k^2,J)}
{\partial
  J_{k_{0}^{'}1}^{(2,2)}({\bf x},{\bf x'})}
\Big|_{J,{\tilde J}=0}\delta _{1k_0}=
\nonumber \\[0.3cm]
&&2(\frac{\Delta k^2}{2}+
i\eta )\int d^d{\bf y}\frac{\partial^2\overline{Z_f}
(k_{0}^{2},\Delta  k^2,J)}
{\partial
  J_{k_{0}^{'}k_{0}}^{(2,1)}({\bf y},{\bf y})\partial
  J_{11}^{(1,2)}({\bf x},{\bf x'})}\Big|_{J,{\tilde J}=0}
\nonumber \\[0.3cm]
&&+2i(\frac{\Delta k^2}{2}+
i\eta )\int d^d{\bf y}\frac{\partial^2\overline{Z_f}
(k_{0}^{2},\Delta  k^2,J)}
{\partial
  J_{k_0k_{0}^{'}}^{(1,2)}({\bf y},{\bf y})\partial
  J_{11}^{(1,2)}({\bf x},{\bf x'})}\Big|_{J,{\tilde J}=0}
\nonumber \\[0.3cm]
&&+\Delta k^2
\int d^d{\bf y}\frac{\partial^2\overline{Z_f}
(k_{0}^{2},\Delta  k^2,J)}
{\partial
  {\tilde J}{k_0 k_{0}^{'}}^{(1,2)}({\bf y},{\bf y})
J_{11}^{(1,2)}({\bf x},{\bf x'})}\Big|_{J,{\tilde J}=0}
\nonumber \\[0.3cm]
&&+\Delta k^2
\int d^d{\bf y}\frac{\partial^2\overline{Z_f}
(k_{0}^{2},\Delta  k^2,J)}
{\partial
  {\tilde J}{k_{0}^{'}k_0}^{(2,1)}({\bf y},{\bf y})
J_{11}^{(1,2)}({\bf x},{\bf x'})}\Big|_{J,{\tilde J}=0} \ .
\label{ward} 
\end{eqnarray}
Replacing $\overline{Z_f}$ everywhere by $Z$, using definition of $Z$ 
and putting $k_0=k_{0}^{'}=1$ we immediately obtain a Ward identity in
the form of Eq. (\ref{wac}). Indeed, Eq. (\ref{tre}) corresponds to the 
l.h.s. of Eq. (\ref{wac}), whereas Eqs. (\ref{pyt},\ref{sem}) lead to the 
r.h.s. of Eq. (\ref{wac}).

\section{Conclusions}
\label{con}

We have derived a non--linear supersymmetric ${\bf \sigma}$--model for
classical scalar waves. We have shown that in the weak disorder limit
($k_0\ell\gg 1$), the effective Lgrangean of this model is identical
to the one for electrons. In this limit, the main difference between
the wave equations for classical and Schr\"odinger waves, the
frequency dependence of the random potential, does not lead to
different wave behavior. We have also shown that the Ward identities
for classical and for Schr\"odinger waves are both fulfilled by the
same effective Lagrangean. This is due to the different source terms.
Outside the regime of weak disorder, i.e. at low frequencies, the
frequeny dependence of the disorder potential for classical waves
suppresses disorder effects altogether. This does not happen for
electrons. This low--frequency domain is not accessible to the
non--linear ${\bf \sigma}$--model.

\acknowledgements{V. K. appreciates very useful discussions with 
Dr. A. M\"uller-Groeling and Dr. Y. Fyodorov.
V. K. gratefully acknowledges the support of 
a MINERVA Fellowship. 
B. E. wishes to thank Dr. A. M\"uller-Groeling and
Prof. A. Nourreddine 
for valuable discussions.}


\begin{references}

\bibitem[*]{brahim}Also at Universit\'e Choua\"\i b 
Doukkali,  Facult\'e des Sciences, El Jadida, 
Morocco.

\bibitem{review}  Y. Imry, {\it Introduction to Mesoscopic Physics}, 
(New York Oxford, Oxford University Press, 1997).
\bibitem{reviewlight} For a review, see {\it Scattering and
  Localization of Classical Waves in Random Media}, edited by Ping
  Sheng (World Scientific, Singapore, 1990); {\it Photonic Band Gaps 
and Localization}, edited by C. M. Sokoulis (Plenum, New York, 1993);
\bibitem{Efe83}K. B. Efetov, Adv. Phys. {\bf 32} (1983) 53.
\bibitem{john} S. John and M. J. Stephen, Phys. Rev. B {\bf 28}, 
6358 (1983);
\bibitem{Ver85}J. J. M. Verbaarschot, M. R. Zirnbauer, and
H. A. Weidenm\"uller, Phys. Rep. {\bf 129} (1985) 387. 




\end{references}
\end{document}